\documentclass[aps,print,twocolumn,superscriptaddress,showpacs]{revtex4}
\usepackage{amsfonts}
\usepackage{amsmath}
\usepackage{amssymb}
\usepackage{graphicx}

\begin{document}

\title{Pairing imbalance in BCS-BEC crossover of inhomogeneous
three-component Fermi-gas in two dimensions}
\author{Jiajia Du}
\affiliation{Institute of Theoretical Physics, Shanxi University, Taiyuan 030006, China}
\author{Junjun Liang}
\affiliation{Department of Physics, Shanxi University, Taiyuan 030006, China}
\author{J.-Q. Liang}
\thanks{Email address: jqliang@sxu.edu.cn}
\affiliation{Institute of Theoretical Physics, Shanxi University, Taiyuan 030006, China}
\pacs{03.75.Hh, 03.75.Ss, 05.30.Fk, 74.20.Fg}

\begin{abstract}
We in this paper investigate the phase diagram associated with the BCS-BEC
crossover of a three-component ultracold superfluid-Fermi-gas of different
chemical-potentials and equal masses in two dimensions. The gap order
parameter and number densities are found analytically by using the
functional path-integral method. The balance of paring will be broken in the
free space due to the unequal chemical-potentials. We obtain the same
particle number-density and condensed fraction in the BCS superfluid phase
as that in a recent paper (Phys. Rev. A 83, 033630 ), while the Sarma phase
of coexistence of normal and superfluid Fermi gases is the characteristics
of inhomogeneous system. The minimum ratio of BCS superfluid phase becomes
1/3 in the BCS limit corresponding to the zero-ratio in the two-component
system in which the critical point of phase separation is $\epsilon
_{B}/\epsilon _{F}=2$ but becomes 3 in the three-component case.
\end{abstract}

\volumenumber{}
\issuenumber{}
\eid{}
\date{}
\received[Received text]{}
\revised[Revised text]{}
\accepted[Accepted text]{}
\published[Published text]{}
\startpage{1}
\endpage{}
\maketitle

\section{Introduction}

Physics related to the crossover from Bardeen-Cooper-Schrieffer (BCS)
superfluid of Fermi-pairs to the Bose-Einstein condensate (BEC) of molecular
dimers in ultracold Fermi gas (FG) has become a central topic in both
experimental\cite{S. Jochim}\cite{M. Greiner}\cite{Zwierlein1}\cite%
{Zwierlein2} and theoretical\cite{Bloch} communities. The Feshbach resonance
plays an important role in practical experiments to manipulate the
interaction between particles in a wide range, which is crucial to form the
Fermi-pairs and induce the crossover.

Researches on the population balance and imbalance in the crossover for
nonuniform (the unequal chemical potentials or masses) two-component FG in
2-dimension (2D)\cite{He22}\cite{G. J. Conduit}\cite{M. Marini}\cite{Jiajia
Du}\cite{Meera} and 3-dimension (3D) \cite{Iskin}\cite{W. Vincent}\cite{C.
A. Regal}\cite{Hui Hu}\cite{Wang Chang Su}, have attracted considerable
attentions in recent years since\textbf{\ }unequal particle-number in
different components may lead to the mismatch of Fermi surfaces, where the
different components refer usually to atoms of different internal states or
different isotopes. It has been demonstrated that there exist various
phenomena, such as the Sarma superfluidity\cite{G. Sarma} with unmatched
Fermi surfaces and phase separation between paired-superfluid and
excess-normal fermions depending on the population imbalance\textbf{\ }in
the 3D resulted from the unequal chemical-potentials and masses.\textbf{\ }%
The Sarma phase stands for a shell structure of the FG in which the core
region is superfluid, while the outer region is normal\cite{K. B. Gubbels}.
The phase separation refers generally to the spatial separation of unpaired
fermions from a BCS superfluid of equal densities because of the existence
of pairing gap in the superfluid region\cite{Y.Shin}. Thus the Sarma phase
is also a type of phase separation.

2D FG possess a great advantage that the variation of bound-state energy in
2D is more suitable to display the BCS-BEC crossover than the interaction
strength in the 3D, where a critical coupling strength is required to form a
s-wave bound state\cite{Miyake K. Miyake}\cite{Mohit Randeria}\textbf{. }%
Both the energy-gap order-parameter, condensate density and population
imbalance can be expressed as a function of binding energy and the phase
boundary between\textbf{\ }normal and Sarma phases is identified in the
BCS-BEC crossover\cite{Jiajia Du}

The three-component FG have been investigated in 2D\cite{Theja N. De}\cite%
{Luca1}\cite{Luca2} and 3D\cite{T. Paananen}\cite{A. N. Wenz3}\cite{J. P.
Martikainen}\cite{G. Catelani} free-spaces respectively showing the
asymmetry of phase diagram by the unequal interaction lengths\cite{G.
Catelani}. However, the asymmetry is not only induced by the unequal
physical quantities but also by the pairing itself between different
components even in a homogeneous system, which leads to population
imbalance. It is demonstrated that for a homogeneous\textbf{\ }%
three-component ultracold FG with a $U(3)$-invariant attractive interaction,
the system exhibits population imbalance in the crossover, which connects
the BCS regime of pairing state and the BEC limit with three species of
molecules, since the paring breaks $U(3)$ symmetry\textbf{. }The spontaneous
population imbalance, first noted in Ref. \cite{Lianyi} is an important
feature of the three-component system. Moreover it is shown in Refs. \cite%
{R. W. Cherng} and \cite{Tomoki} that BCS superfluidity and population
imbalance can coexist in weak-coupling regime of multicomponent Fermi
systems.\textbf{\ }The BCS-BEC crossover of uniform three-component FG with $%
U(3)$ symmetry was investigated recently\cite{Tomoki}\textbf{\ }%
demonstrating the asymmetry induced by paring. The three-component
superfluidity has been realized already by the lowest three hyperfine
spin-states in the mixture of $^{6}$Li atoms and, moreover,\textbf{\ }the
experimental analysis of the condensate fraction of ultracold two-component
FG in the BCS-BEC crossover is shown to be in good agreement with the
mean-field theoretical predictions\cite{Luca3}\cite{G. Ortiz} and Monte
Carlo simulations\cite{G. E. Astrakharchik}.

In a most recent paper\cite{Luca1}\cite{Luca2} the analytical formulas of
number densities and condensate fraction for homogeneous three-component FG
of both 3D and 2D are obtained in the BCS-BEC crossover in which, although
the third component is assumed not to have interaction with the other two
components the asymmetric interactions affect the formation of condensation
and lead to the population imbalance\cite{Luca1}\cite{Luca2}.

We in this paper investigate the BCS-BEC crossover of the three-component FG
with asymmetric interactions the same as in Ref. \cite{Luca1} and Ref. \cite%
{Luca2}, but unequal chemical potentials in 2D free-space. We are
particularly interested in the Sarma superfluidity and phase separation in
the crossover induced by the asymmetry of two interacting components.
Analytical results of the particle-number density and the gap order
parameter are obtained with the functional path-integral method. Then we
analyze the condensate fraction based on the phase diagram and demonstrate
that the Sarma phase is the characteristics of the asymmetric system.

\section{ Particle-number and gap equations}

The non-uniform three-component FG, which we are going to consider, can be
described by the Hamiltonian $\left( \hbar =1\right) $%
\begin{align}
H& =\sum_{\mathbf{k},i=\mathbf{R,G,B}}\xi _{\mathbf{k,}i}\psi _{\mathbf{k,}%
i}^{\dagger }\psi _{\mathbf{k,}i}  \notag \\
& +\!\!\!\sum_{\mathbf{k,k}^{/},\mathbf{q}}(g_{R,G}\psi _{\mathbf{k}+\mathbf{%
q}/2,R}^{\dagger }\psi _{-\mathbf{k}+\mathbf{q}/2,G}^{\dagger }\psi _{-%
\mathbf{k}^{/}+\mathbf{q}/2,G}\psi _{\mathbf{k}^{/}+\mathbf{q}/2,R}  \notag
\\
& +g_{B,R}\psi _{\mathbf{k}+\mathbf{q}/2,R}^{\dagger }\psi _{-\mathbf{k}+%
\mathbf{q}/2,B}^{\dagger }\psi _{-\mathbf{k}^{/}+\mathbf{q}/2,B}\psi _{%
\mathbf{k}^{/}+\mathbf{q}/2,R}  \notag \\
& +g_{B,G}\psi _{\mathbf{k}+\mathbf{q}/2,B}^{\dagger }\psi _{-\mathbf{k}+%
\mathbf{q}/2,G}^{\dagger }\psi _{-\mathbf{k}^{/}+\mathbf{q}/2,G}\psi _{%
\mathbf{k}^{/}+\mathbf{q}/2,B})
\end{align}%
where $\psi _{\mathbf{k,}i}^{\dagger }$\ ($\psi _{\mathbf{k,}i}$) is the
Fermi-field creation (annihilation) operator creating ( annihilating) a
particle of color-index $i=R,G,B\ $(denoting red, green and blue
respectively) and momentum $k$. $\xi _{\mathbf{k},i}=\varepsilon _{\mathbf{k}%
,i}-\mu _{\mathbf{k},i}$ with $\varepsilon _{\mathbf{k}}=k^{2}/\left(
2m\right) $\ being the kinetic energy. The three components possess the same
mass $m$ but unequal chemical potentials $\mu _{i}.$\ $g$\ with a negative
value is used to describe the attractive interatomic interaction. The rest
three sums in the Hamiltonian denote the pairing among particles of
different colors while there is no pairing in the same component of FG. We
furthermore assume that the pairing takes place only between two components,
say, the red and green particles. Thus the Hamiltonian is reduced to a
simple form%
\begin{align}
H& =\sum_{\mathbf{k},i=\mathbf{R,G,B}}\xi _{\mathbf{k,}i}\psi _{\mathbf{k,}%
i}^{\dagger }\psi _{\mathbf{k,}i}  \notag \\
& +\!\!\sum_{\mathbf{k,k}^{/},\mathbf{q}}g_{R,G}\psi _{\mathbf{k}+\mathbf{q}%
/2,R}^{\dagger }\psi _{-\!\mathbf{k}+\mathbf{q}/2,G}^{\dagger }\psi _{\!-\!%
\mathbf{k}^{/}+\mathbf{q}/2,G}\psi _{\mathbf{k}^{/}+\mathbf{q}/2,R}
\label{hamiltonian}
\end{align}%
By using the functional path-integral treatment, we obtain the
thermodynamical potential of saddle-point, which can be written as\cite{E.
H. Vivas C.}%
\begin{equation}
\Omega _{0}\left( \Delta ,\mu _{R},\mu _{G},\mu _{B}\right) \!=\!\sum_{%
\mathbf{k}}[f(E_{\mathbf{k},+})E_{\mathbf{k},+}\!+\!f(E_{\mathbf{k},-})E_{%
\mathbf{k},-}  \notag
\end{equation}%
\begin{equation}
\!+\!\xi _{\mathbf{k},G}\!-\!E_{\mathbf{k},-}\!+\!f_{F}\left( \mu
_{B}\right) \xi _{\mathbf{k,}B}\!-\!\frac{\Delta ^{2}}{g}]
\label{thermodynamical}
\end{equation}%
where $E_{\mathbf{k},\pm }=(\xi _{\mathbf{k},+}^{2}+\Delta ^{2})^{1/2}\pm
\xi _{\mathbf{k},-}$\ are the quasiparticle and the quasihole energies with $%
\xi _{\mathbf{k},\pm }=(\xi _{\mathbf{k},R}\pm \xi _{\mathbf{k}%
,G})/2=k^{2}/(2m)-\mu _{\pm }$ and energy-gap $\Delta $ is the
order-parameter\textbf{\ }of FG. The reexpressed chemical potentials $\mu
_{\pm }=(\mu _{R}\pm \mu _{G})/2$ indicate the chemical potential imbalance
between red and green particles. The particle number of color-index $i$ is
obtained from the saddle point approximation as $N_{0,i}=-\partial \Omega
_{0}\left( \Delta ,\mu _{R},\mu _{G},\mu _{B}\right) /\partial \mu _{i}$,
and the explicit formulas of the three-color particle-number densities $%
n_{i}=N_{0,i}/V$ ($i=R,G,B$) are given respectively by \textbf{\ }%
\begin{equation}
n_{R}=\int \frac{d^{2}\mathbf{k}}{\left( 2\pi \right) ^{2}}\left[ \left(
1-f(E_{\mathbf{k},-})\right) v_{\mathbf{k}}^{2}+f(E_{\mathbf{k},+})u_{%
\mathbf{k}}^{2}\right]   \label{num1}
\end{equation}%
\begin{align}
n_{G}& =\int \frac{d^{2}\mathbf{k}}{\left( 2\pi \right) ^{2}}\left[ \left(
1-f(E_{\mathbf{k},+})\right) v_{\mathbf{k}}^{2}+f(E_{\mathbf{k},-})u_{%
\mathbf{k}}^{2}\right]   \label{num2} \\
n_{B}& =\int \frac{d^{2}\mathbf{k}}{\left( 2\pi \right) ^{2}}\left[
f_{F}\left( \mu _{B}\right) \right]   \label{num3}
\end{align}%
\textbf{\ }where\textbf{\ }$u_{\mathbf{k}}^{2}=\left( 1+\xi _{\mathbf{k}%
,+}/E_{\mathbf{k},+}\right) /2$\textbf{, }$v_{\mathbf{k}}^{2}$\textbf{\ }$%
=\left( 1-\xi _{\mathbf{k},+}/E_{\mathbf{k},+}\right) /2$\textbf{\ }and the
Fermi thermal distribution-function has the usual form\textbf{\ }$%
f(x)=\left( \exp (\beta x)+1\right) ^{-1}$\textbf{. }The number equations (%
\ref{num1}) and (\ref{num2}) of paired components must be coupled with the
gap equation given by $\delta S_{0}/\delta \Delta =0,$ ( $S_{0}=\beta \Omega
_{0}$), which is
\begin{equation}
-\frac{1}{g}=\sum_{\mathbf{k}}\frac{1-f(E_{\mathbf{k},+})-f(E_{\mathbf{k},-})%
}{2\sqrt{\xi _{\mathbf{k},+}^{2}+\Delta ^{2}}}.  \label{Gap0}
\end{equation}%
The binding energy $\epsilon _{B}$, which is more suitable than interaction
strength $g$ to describe BCS-BEC crossover as demonstrated in the
introduction, can be determined from the bound-state energy equation
\begin{equation}
-\frac{1}{g}=\frac{1}{\Omega }\sum_{\mathbf{k}}\frac{1}{\frac{k^{2}}{m}%
+\epsilon _{B}}.  \label{bound0}
\end{equation}%
The particle-number (\ref{num1}), (\ref{num2}), (\ref{num3})and gap equation
(\ref{Gap0}) are going to be solved self-consistently under the
zero-temperature limit that $\theta (-E_{\mathbf{k,\pm }})=\lim_{\beta
\rightarrow \infty }f(E_{\mathbf{k},\pm })$, where $\theta (x)$\ is the
Heaviside function and the analytic results are shown in the following
section.

\section{Criterion of the phase transition}

At zero-temperature the saddle-point self-consistent equations (\ref{num1}),
(\ref{num2}), (\ref{num3}) and (\ref{Gap0}) are suitable to describe the
BCS-BEC crossover, from which we can calculate the particle number and the
binding energy $\epsilon _{B}$. When the quasiparticle energy vanishes $E_{%
\mathbf{k},+}=0$, the zero-point energies $\varepsilon _{\pm }$ are found as%
\begin{align}
\varepsilon _{\pm }& =\frac{1}{2}\left[ \left( \mu _{R}+\mu _{G}\right) \pm
\sqrt{\left( \mu _{R}-\mu _{G}\right) ^{2}-4\Delta ^{2}}\right]  \notag \\
& =\mu _{+}\pm \sqrt{\mu _{-}^{2}-\Delta ^{2}}  \label{zero-point energy}
\end{align}%
where we have used the property of Fermi thermal distribution-function that $%
f(x)=0$ at $x<0$ and $f(x)=1$ at $x>0$ at zero-temperature with the integral
variable, i.e. the kinetic energy $\varepsilon _{\mathbf{k}}=k^{2}/\left(
2m\right) $. In the chemical potential range $-\Delta <\mu _{-}<\Delta ,$
the zero-point energies $\varepsilon _{\pm }$ are complex values, so the
quasiparticle energies $E_{\mathbf{k},+}$ and $E_{\mathbf{k},-}$ are
positive indicating the BCS superfluid state. When $\left\vert \mu
_{-}\right\vert >\Delta $, $\varepsilon _{\pm }$ are real and moreover in
the branch of $\varepsilon _{\pm }>0,$ the quasienergy $E_{\mathbf{k},+}$ is
negative, while the kinetic energy in the region $\varepsilon
_{-}<\varepsilon _{\mathbf{k}}<\varepsilon _{+}$ the quasienergy $E_{\mathbf{%
k},-}$ is always positive. To have the kinetic energy $\varepsilon _{\mathbf{%
k}}$ being positive, its value-range should be restricted to $\left[
0,\varepsilon _{+}\right] $ when $\varepsilon _{-}<0$ giving rise to the
Sarma phase, which\textbf{\ }is the coexistence of the normal and superfluid
FG.

The particle-number of BEC phase is strictly related to the off-diagonal
long-range order (ODLRO) in interacting many-body systems, which was
proposed in terms of the asymptotic behavior of two-body density matrix\cite%
{Yang} defined as $\rho (\mathbf{r}_{R}^{^{\prime }},\mathbf{r}%
_{G}^{^{\prime }},\mathbf{r}_{R},\mathbf{r}_{G})\equiv \left\langle \psi
_{R}^{\dagger }(\mathbf{r}_{R}^{^{\prime }})\psi _{G}^{\dagger }(\mathbf{r}%
_{G}^{^{\prime }})\psi _{R}(\mathbf{r}_{R})\psi _{G}(\mathbf{r}%
_{G})\right\rangle ,$ where $\psi _{i}(\mathbf{r})$ and $\psi _{i}^{\dagger
}(\mathbf{r})$ are the fermionic creation and annihilation operators of $i$%
-th species in the position $\mathbf{r}$ respectively and $\left\langle
...\right\rangle $ stands for quantum-mechanical expectation value. For a
homogeneous FG, the ODLRO of Fermi superfluid is characterized by the
asymptotic behavior of density matrix $\rho (\mathbf{r}_{R}^{^{\prime }},%
\mathbf{r}_{G}^{^{\prime }},\mathbf{r}_{R},\mathbf{r}_{G})$ for $\left\vert
\mathbf{r}_{R}-\mathbf{r}_{R}^{^{\prime }}\right\vert ,$ $\left\vert \mathbf{%
r}_{G}-\mathbf{r}_{G}^{^{\prime }}\right\vert \rightarrow \infty .$ While
the largest eigenvalue $N_{c}$ of the two-body density matrix gives rise to
the number of Fermi pairs in BEC phase, which is denoted by%
\begin{equation*}
N_{c}=\int d^{2}\mathbf{r}_{R}d^{2}\mathbf{r}_{G}\left\vert \left\langle
\psi _{R}(\mathbf{r}_{R})\psi _{G}(\mathbf{r}_{G})\right\rangle \right\vert
^{2}
\end{equation*}%
and it is straightforward to show\cite{Campbell} that
\begin{equation}
N_{c}=\int \frac{d^{2}\mathbf{k}}{(2\pi )^{2}}u_{k}^{2}v_{k}^{2}  \label{nc}
\end{equation}%
In this paper, we moreover define a physical quantity called ratio of
unpaired particle-number imbalance as:
\begin{equation}
P=\frac{n_{R}-n_{G}+n_{B}}{n_{R}+n_{G}+n_{B}}  \label{ratio}
\end{equation}%
which with value in the range $\left[ 0,1\right] $ can identify the phase
transition.

\subsection{ Sarma superfluid phase}

At first, we consider the zero-point energy range $\left[ \varepsilon
_{-},\varepsilon _{+}\right] $, which can be classified into three cases
that $0<\varepsilon _{-}<\varepsilon _{+},$ $\varepsilon _{-}<0<\varepsilon
_{+},$ and $\varepsilon _{-}<\varepsilon _{+}<0$. The third case, which is
the same as in the chemical potential region $-\Delta <\mu _{-}<\Delta $, is
the BCS phase. The two other cases are both the Sarma phase. When\textbf{\ }$%
E_{\mathbf{k},+}<0$ with $k\neq 0$ and $0<\varepsilon _{-}<\varepsilon _{+},$
the Sarma phase is in the BCS regime while in the BEC regime for $k=0$, $%
\varepsilon _{-}<0<\varepsilon _{+}$. The particle-number density equation
is
\begin{equation}
n=n_{R}+n_{G}+n_{B}=m\epsilon _{F}/\pi   \label{total number}
\end{equation}%
where $\epsilon _{F}$ is the Fermi energy. The number-density difference
between red and green particles of the Eqs. (\ref{num1}) and (\ref{num2}) is
seen to be
\begin{equation}
n_{R}-n_{G}=m/\left( 2\pi \right) \left( \varepsilon _{+}-\varepsilon
_{-}\right)   \label{the difference number}
\end{equation}%
Thus particle-number densities in the range $\left[ \varepsilon
_{-},\varepsilon _{+}\right] $ are found from Eqs. (\ref{num1}) and (\ref%
{num2}) as

\begin{align*}
n_{G} & =0 \\
n_{R} & =m/\left( 2\pi\right) \left( \varepsilon_{+}-\varepsilon _{-}\right)
\\
n_{B} & =m\mu_{B}/\left( 2\pi\right) \theta\left( \mu_{B}\right)
\end{align*}
along with vanishing gap $\Delta=0$. For the sake of simplicity, we let the
chemical potential of blue component be $\mu_{B}=\mu_{+}$ throughout the
paper.{}

\subsubsection{Sarma phase $P2$ in the BCS regime}

For the case one $\left( 0<\varepsilon _{-}<\varepsilon _{+}\right) $ with $%
E_{\mathbf{k},+}<0$ , $E_{\mathbf{k},-}>0,$ we have $f_{F}\left( E_{\mathbf{k%
},+}\right) =1$ , $f_{F}\left( E_{\mathbf{k},-}\right) =0$ the particle
number density of green component evaluated from the Eq. (\ref{num2}) is
given by :
\begin{equation}
n_{G}=\left( \int_{0}^{\varepsilon _{-}}+\int_{\varepsilon _{+}}^{\infty
}\right) \tfrac{md\varepsilon }{4\pi }\left( 1/2-\tfrac{\left( \varepsilon
-\mu _{+}\right) }{2\sqrt{\left( \varepsilon -\mu _{+}\right) ^{2}+\Delta
^{2}}}\right)   \label{1,ng}
\end{equation}%
With the result of Eq. (\ref{num1}) together we have%
\begin{equation}
n_{R}-n_{G}=0  \label{1,difference}
\end{equation}%
The energy-gap and bound-state energy are determined by the equations
\begin{equation}
-\frac{1}{g}=\left( \int_{0}^{\varepsilon _{-}}+\int_{\varepsilon
_{+}}^{\infty }\right) \frac{md\varepsilon }{4\pi }\frac{1}{\sqrt{\left(
\varepsilon -\mu _{+}\right) ^{2}+\Delta ^{2}}},  \label{1,energy}
\end{equation}%
\begin{equation}
-\frac{1}{g}=\frac{1}{2}\int_{0}^{\infty }\frac{md\varepsilon }{4\pi }\frac{1%
}{\varepsilon +\epsilon _{B}}  \label{1,bound state}
\end{equation}%
which can be combined together as%
\begin{align}
& \left( \int_{0}^{\varepsilon _{-}}+\int_{\varepsilon _{+}}^{\infty
}\right) \frac{md\varepsilon }{4\pi }\frac{1}{\sqrt{\left( \varepsilon -\mu
_{+}\right) ^{2}+\Delta ^{2}}}  \notag \\
& =\frac{1}{2}\int_{0}^{\infty }\frac{md\varepsilon }{4\pi }\frac{1}{%
\varepsilon +\epsilon _{B}}  \label{together}
\end{align}%
where $\omega _{\pm }=\frac{\varepsilon _{\pm }}{\Delta },x_{0}=\frac{\mu
_{+}}{\Delta },x^{2}=\frac{\varepsilon }{\Delta }$. Working out the
integration in Eq. (\ref{1,ng}), the particle number density has the form:%
\begin{align}
n_{G}\!\!& =\!\!\frac{m\Delta }{4\pi }[\omega _{-}-\omega _{+}\!\!+\!\!\sqrt{%
\left( \omega _{+}\!\!-\!\!x_{0}\right) ^{2}\!\!+\!\!1}\!\!  \notag \\
& -\!\!\sqrt{\left( \omega _{-}\!\!-\!\!x_{0}\right) ^{2}\!\!+\!\!1}%
\!\!+\!\!x_{0}\!\!+\!\!\sqrt{x_{0}^{2}\!\!+\!\!1}]  \label{1,ng result}
\end{align}%
\begin{equation}
n_{B}=m\mu _{+}/\left( 2\pi \right) \theta \left( \mu _{+}\right)
\label{1,nb result}
\end{equation}%
and the gap parameter is obtained from Eq. (\ref{together})
\begin{equation*}
\frac{\epsilon _{B}}{\Delta }=\frac{\left[ -x_{0}\!\!+\!\!\sqrt{%
x_{0}^{2}\!\!+\!\!1}\right] \left[ \left( \omega _{+}\!\!-\!\!x_{0}\right)
\!\!+\!\!\sqrt{\left( \omega _{+}\!\!-\!\!x_{0}\right) ^{2}\!\!+\!\!1}\right]
}{\left[ \left( \omega _{-}\!\!-\!\!x_{0}\right) \!\!+\!\!\sqrt{\left(
\omega _{-}\!\!-\!\!x_{0}\right) ^{2}\!\!+\!\!1}\right] }
\end{equation*}%
it can be easily rewritten as%
\begin{equation}
\frac{\Delta }{\epsilon _{F}}=\frac{3}{\sqrt{\left( \omega
_{+}\!\!-\!\!x_{0}\right) ^{2}\!\!+\!\!1}\!\!-\!\!\sqrt{\left( \omega
_{-}\!\!-\!\!x_{0}\right) ^{2}\!\!+\!\!1}\!\!+\!\!x_{0}\theta
(x_{0})\!\!+\!\!x_{0}\!\!+\!\!\sqrt{x_{0}^{2}\!\!+\!\!1}}
\label{1,gap energy}
\end{equation}%
For $n=n_{R}+n_{G}+n_{B}=m\epsilon _{F}/\pi $ with $k_{F}=\sqrt{\frac{4\pi n%
}{3}},$ the binding energy $\epsilon _{B}$ is seen to be%
\begin{widetext}
\begin{equation}
\frac{\epsilon_{B}}{\epsilon_{F}}\!=\!\frac{3\left[  -x_{0}\!+\!\sqrt{x_{0}^{2}%
\!+\!1}\right]  }{\left[  \left(  \omega_{-}\!-\!x_{0}\right)  \!+\!\sqrt{\left(
\omega_{-}\!-\!x_{0}\right)  ^{2}\!+\!1}\right]  }\frac{\left[  \left(  \omega
_{+}\!-\!x_{0}\right)  \!+\!\sqrt{\left(  \omega_{+}\!-\!x_{0}\right)  ^{2}\!+\!1}\right]
}{\left[  \sqrt{\left(  \omega_{+}\!-\!x_{0}\right)  ^{2}\!+\!1}\!-\!\sqrt{\left(
\omega_{-}\!-\!x_{0}\right)  ^{2}\!+\!1}\!+\!x_{0}\theta(x_{0})\!+\!x_{0}\!+\!\sqrt{x_{0}^{2}%
+1}\right]  }
\label{1,binding energy}%
\end{equation}
\end{widetext}and the ratio of unpaired particle-number imbalance $P$ is
easily obtained from Eq. (\ref{ratio}) as:%
\begin{equation}
P\!\!=\!\!\frac{\omega _{+}\!\!-\!\!\omega _{-}\!\!+\!\!x_{0}\theta (x_{0})}{%
\sqrt{\left( \omega _{+}\!\!-\!\!x_{0}\right) ^{2}\!\!+\!\!1}\!\!-\!\!\sqrt{%
\left( \omega _{-}\!\!-\!\!x_{0}\right) ^{2}\!\!+\!\!1}\!\!+\!\!x_{0}\theta
(x_{0})\!\!+\!\!x_{0}\!\!+\!\!\sqrt{x_{0}^{2}\!\!+\!\!1}}  \label{1,p}
\end{equation}%
The condensate fraction $n_{c}\!\!=N_{c}/V$ is found from the Eq. (\ref{nc})
as%
\begin{equation}
n_{c}\!\!=\!\!\frac{\Delta m}{8\pi }\left[ \frac{\pi }{2}\!\!+\!\!\arctan
\left( \frac{\mu _{+}}{\Delta }\right) \!\!-\!\!2\arctan \left( \sqrt{\left(
\frac{\mu _{-}}{\Delta }\right) ^{2}\!\!-\!\!1}\right) \right]   \label{1,n0}
\end{equation}%
{}

\subsubsection{Sarma phase $P1$ in BEC regime}

In case two $\left( \varepsilon_{-}<0<\varepsilon_{+}\right) $ the integral
range is from $0$ to $\varepsilon_{+}$ where the minimum value of $E_{%
\mathbf{k,+}}$ is located at $k=0$, so the system is in the BEC regime. The
number density and energy-gap can be shown from the Eqs. (\ref{the
difference number}), (\ref{num2}), (\ref{Gap0}) and (\ref{bound0})as:%
\begin{equation}
n_{R}-n_{G}=\frac{m}{2\pi}\varepsilon_{+}  \label{2.difference}
\end{equation}%
\begin{equation}
n_{G}=\frac{m\Delta}{4\pi}\left[ \sqrt{\left( \omega_{+}-x_{0}\right) ^{2}+1}%
-\omega_{+}+x_{0}\right]  \label{2,ng}
\end{equation}

\begin{equation}
\frac{\Delta }{\epsilon _{F}}=\frac{3}{\sqrt{\left( \omega _{+}-x_{0}\right)
^{2}+1}+x_{0}\theta (x_{0})+x_{0}}  \label{2,gap energy}
\end{equation}%
The binding energy is%
\begin{equation}
\frac{\epsilon _{B}}{\epsilon _{F}}=\frac{3\left[ \left( \omega
_{+}-x_{0}\right) +\sqrt{\left( \omega _{+}-x_{0}\right) ^{2}+1}\right] }{%
\sqrt{\left( \omega _{+}-x_{0}\right) ^{2}+1}+x_{0}\theta (x_{0})+x_{0}}
\label{2,bound}
\end{equation}%
and
\begin{equation}
P=\frac{\omega _{+}+x_{0}\theta (x_{0})}{\sqrt{\left( \omega
_{+}-x_{0}\right) ^{2}+1}+x_{0}\theta (x_{0})+x_{0}}  \label{2,p}
\end{equation}%
We can easily get the condensate fraction
\begin{equation}
n_{c}=\frac{\Delta m}{8\pi }\left[ \frac{\pi }{2}-\arctan \left( \sqrt{%
\left( \frac{\mu _{-}}{\Delta }\right) ^{2}-1}\right) \right]  \label{2,nc}
\end{equation}%
{}

\subsection{BCS superfluid phase}

For positive $E_{\mathbf{k,\pm }}$ and $\varepsilon _{-}<\varepsilon _{+}<0,$
particle number densities of the red and green components are equal seen
from the Eqs. (\ref{num1}), (\ref{num2}) and the chemical potential is in
the region $-\Delta <\mu _{-}<\Delta $. Number densities and gap parameter
are given by:%
\begin{equation}
n_{R}=n_{G}=\frac{m\Delta }{4\pi }\left[ x_{0}+\sqrt{x_{0}^{2}+1}\right]
\label{3,ng}
\end{equation}%
\begin{equation}
\frac{\Delta }{\epsilon _{F}}=\frac{3}{x_{0}\theta (x_{0})+x_{0}+\sqrt{%
x_{0}^{2}+1}}  \label{3,gap energy}
\end{equation}%
In this case, the binding energy can be shown as Eq. (\ref{1,binding energy})%
\begin{equation}
\frac{\epsilon _{B}}{\epsilon _{F}}=\frac{3\left[ \sqrt{x_{0}^{2}+1}-x_{0}%
\right] }{x_{0}\theta (x_{0})+x_{0}+\sqrt{x_{0}^{2}+1}}  \label{3,bound}
\end{equation}%
and
\begin{equation}
P=\frac{x_{0}\theta (x_{0})}{x_{0}\theta (x_{0})+x_{0}+\sqrt{x_{0}^{2}+1}}
\label{3,p}
\end{equation}%
\begin{figure}[tbp]
\begin{center}
\includegraphics[
height=2.0in,
width=4.0in
]{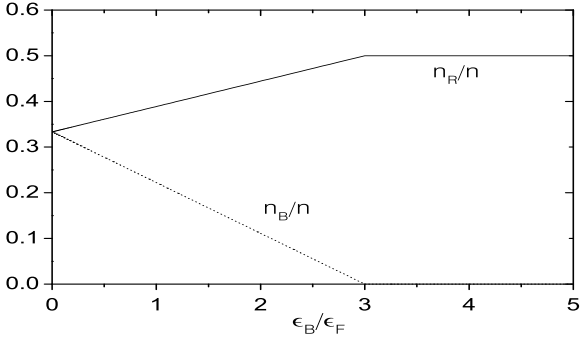}
\end{center}
\caption{(Colour online) The number-density ratios $n_{R}/n$, and $n_{B}/n$,
as a function of the bound-state energy $\protect\epsilon _{B}/\protect%
\epsilon _{F}.$}
\end{figure}
The ratios of particle number densities obtained from Eqs. (\ref{3,ng}) and (%
\ref{1,nb result}) are given by
\begin{equation}
\frac{n_{R}}{n}=\frac{n_{G}}{n}=\frac{1}{2}\frac{x_{0}+\sqrt{x_{0}^{2}+1}}{%
x_{0}\theta (x_{0})+x_{0}+\sqrt{x_{0}^{2}+1}}  \label{fraction nr}
\end{equation}%
and
\begin{equation}
\frac{n_{B}}{n}=1-\frac{x_{0}+\sqrt{x_{0}^{2}+1}}{x_{0}\theta (x_{0})+x_{0}+%
\sqrt{x_{0}^{2}+1}}  \label{fraction nb}
\end{equation}%
which in agreement with the Ref. \cite{Luca1} are displayed in Fig.1, where
we see the balance of number densities $n_{R}/n=n_{G}/n=n_{B}/n=1/3$ .in the
BCS regime. With increase of bound-state energy $\epsilon _{B}/\epsilon
_{F}, $ the number density of red particles $n_{R}/n$ increases while $%
n_{B}/n$ decreases. At the energy value $\epsilon _{B}/\epsilon _{F}=3,$ the
red density-ratio becomes 1/2 and the blue one is zero. The red and green
density-ratios are able to sustain 1/2 with higher value of $\epsilon
_{B}/\epsilon _{F}$ in the BEC regime. The condensate fraction Eq. (\ref{nc}%
) gives rise to the expression
\begin{equation}
\frac{n_{c}}{n}=\frac{1}{4}\frac{\left[ \frac{\pi }{2}+\arctan \left(
x_{0}\right) \right] }{x_{0}\theta (x_{0})+x_{0}+\sqrt{x_{0}^{2}+1}}
\label{condensate fraction}
\end{equation}%
{}

\subsection{ Phase transition and boundary}

The frequency $\omega_{\pm}-x_{0}=\pm\sqrt{y_{0}^{2}-1}$ with $y_{0}=\mu
_{-}/\Delta,$ is a characteristic quantity of the system at hand with which
the ratio $P$ and bound state energy $\epsilon_{B}/\epsilon_{F}$ can be
expressed as the function of two parameters $x_{0}$ and $y_{0}.${}

In the Sarma phase ratio $P$ and bound state energy $\epsilon_{B}/\epsilon
_{F}$ become functions of parameters $x_{0}$ and $y_{0}$%
\begin{equation}
P=\frac{x_{0}\theta(x_{0})+2\sqrt{y_{0}^{2}-1}}{x_{0}\theta(x_{0})+x_{0}+%
\sqrt{x_{0}^{2}+1}}  \label{ratio1}
\end{equation}

\begin{equation}
\frac{\epsilon_{B}}{\epsilon_{F}}=\frac{3\left[ \sqrt{x_{0}^{2}+1}-x_{0}%
\right] \left[ \sqrt{y_{0}^{2}-1}+y_{0}\right] }{\left[ x_{0}%
\theta(x_{0})+x_{0}+\sqrt{x_{0}^{2}+1}\right] \left[ y_{0}-\sqrt{y_{0}^{2}-1}%
\right] }  \label{bound1}
\end{equation}
in the region $0<\varepsilon_{-}<\varepsilon_{+}$, and

\begin{equation}
P=\frac{x_{0}\theta(x_{0})+x_{0}+\sqrt{y_{0}^{2}-1}}{x_{0}%
\theta(x_{0})+x_{0}+y_{0}}  \label{ratio2}
\end{equation}%
\begin{equation}
\frac{\epsilon_{B}}{\epsilon_{F}}=\frac{3\left[ \sqrt{y_{0}^{2}-1}+y_{0}%
\right] }{x_{0}\theta(x_{0})+x_{0}+y_{0}}  \label{bound2}
\end{equation}
in the case of $\varepsilon_{-}<0<\varepsilon_{+}$ respectively. The phase
transition and boundary are determined by the number of zeros of
quasienergies $E_{\mathbf{k},\pm}$ in the momentum space. Sarma phase in two
ranges with one and two zeros respectively, which identify the normal gas,
does not exist in the three-component FG of equal chemical potentials. The
BCS superfluid phase in two ranges has no zero, while two negative
zero-point energies correspond to negative quasienergy also different from
the standard BCS.

In the phase diagram of Fig.2 obtained from Equations of $P$ and $\epsilon
_{B}/\epsilon_{F}$, $P1$\textbf{\ }stands for the Sarma phase with effective
Fermi-surfaces formed by the two zero-points of excitation-spectrum $E_{%
\mathbf{k},+}$ and one negative zero-point energy in the condition $-\sqrt{%
y_{0}^{2}-1}<x_{0}<\sqrt{y_{0}^{2}-1}$.
\begin{figure}[ptb]
\begin{center}
\includegraphics[
height=2.0in,
width=4.0in
]{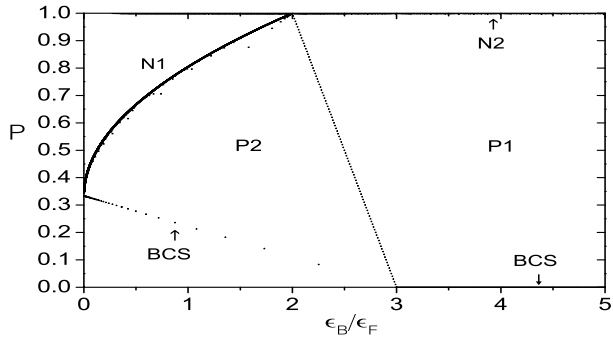}
\end{center}
\caption{(Colour online) Phase diagram in the plane of $\protect\epsilon_{R}/%
\protect\epsilon_{F}$ and $P$. The Sarma phase is divided to two parts $P1$
and $P2$ by the critical line $P=3-\protect\epsilon_{B}/\protect\epsilon_{F}$%
.}
\end{figure}
\ \ \

The Sarma phase $P2$ has effective Fermi-surfaces resulted from two
zero-points of $E_{\mathbf{k},+}$ .in the condition\textbf{\ }$x_{0}>\sqrt{%
y_{0}^{2}-1}$. The BCS phase indicated in Fig 2 has no effective
Fermi-surface of zero quasienergies. More specifically in the condition $%
y_{0}<1$. $E_{\mathbf{k},+}$ has no zero-point in one case, which is
consistent with the classical BCS superconductivity, and two negative
zero-point energies in the other case with the condition $x_{0}<0$ and $%
x_{0}<-\sqrt{y_{0}^{2}-1}.$

From Fig.2, it is found that at the BCS limit ($\epsilon _{B}/\epsilon
_{F}=0 $) $P=1/3$ in both the BCS and the Sarma phases, which is consistent
with the Fig.1 at $\epsilon _{B}/\epsilon _{F}=0$\textbf{\ }and is different
from the two-component FG (in which $P=0$ at the BCS limit) due to the
existence of the third component. The critical point $x_{0}=0,$ where $%
\epsilon _{B}/\epsilon _{F}=3$ and $P=0,$ is a common point of the three
phases. The Sarma phase is divided to two parts $P1$ and $P2$ by the
critical line $P=3-\epsilon _{B}/\epsilon _{F}$, while the two parts are
divided by the line $\epsilon _{B}/\epsilon _{F}=2$ in two-component FG. In
the BCS superfluid phase $P=0$ for $\epsilon _{B}/\epsilon _{F}>3$ and it
decreases from $P=1/3$ to $P=0$ with the increase of $\epsilon _{B}/\epsilon
_{F}$ in the region $\epsilon _{B}/\epsilon _{F}<3$ different from the
two-component FG in which the BCS phase goes along $P=0$ in the $P-\epsilon
_{B}/\epsilon _{F}$ plane. $N1$ and $N2$ in Fig.2 respectively denote the
partially and fully polarized normal phases, which are also abundant phases
and do not exist in the three-component FG of equal chemical-potentials.
With the increase of $P$ and $\epsilon _{B}/\epsilon _{F},$ the phase
transitions of BCS, normal, Sarma phases $P2$, $P1$ take place in order.
When $\epsilon _{B}/\epsilon _{F}>3,$ there are only red and green
components in the crossover and the BCS superfluid phase emerges at $P=0$
i.e. the balance pairing $n_{R}=n_{G},$ in consistence with the Fig.1, which
shows $n_{G}/n=n_{R}/n=1/2.$

For a Fermi gas in 2D, the quantum fluctuations play a more crucial role
with respect to the 3D case, which suppress the formation of long-range
phase coherence at non-zero temperature\cite{SR} and remain non-negligible
even at zero temperature. It has been demonstrated the quantum fluctuations
keep on playing a crucial role also when imbalance is present\cite{TKD}. As
a consequence, in order to describe Fermi gases in the whole range of the
BCS-BEC crossover at zero temperature the mean-field approximation might be
not sufficient, and a more accurate account of quantum fluctuations is
necessary. The research beyond the mean-field approximation is under
development referring to the BCS-BEC crossover.

\section{Conclusions}

The pairing physics in BCS-BEC crossover is studied with particular interest
in the chemical-potential asymmetry induced imbalance. The analytical energy
gap, particle-number density and the condensate fraction of ultracold
three-component FG of unequal chemical-potentials in 2D are shown in exact
agreement with the results in Ref. \cite{Luca1} of homogeneous system, while
Sarma phase, which is divided into two parts $P2$, and $P1$, partially and
fully polarized normal phases\textbf{\ }are new and are resulted from the
asymmetric chemical potentials. Phase diagram and transitions in the $%
P-\epsilon _{B}/\epsilon _{F}$ plane are also compared with the
inhomogeneous two-component FG. Although the third component is assumed not
to pairing with the other two components the ratio $P$, which vanishes in
the BCS limit for the two-component FG\cite{Jiajia Du}, becomes 1/3 in the
three-component system with the critical point of phase separation $\epsilon
_{B}/\epsilon _{F}=3.$ \medskip

\section{Acknowledgment}

One of authors (JJD) is grateful to Dr. Luca Salasnich for helpful
discussions. This work was supported by National Nature Science Foundation
of China (Grant No.11075099).

\end{document}